\documentclass[nohyper,12pt]{JHEP3}
\def\ba{\begin{array}}
\def\ea{\end{array}}
\def\be{\begin{equation}}
\def\ee{\end{equation}}
\def\lbl{\label}
\def \rf{(\ref}

\def\x0{\x_0}
\def\x1{\x_1}

\def\real{{\bf R}}

\def\cd{{\cal {D}}}
\def\cg{{\cal {G}}}
\def\tcg{{\tilde{\cal G}}}

\newtheorem{theorem}{Theorem}
\title{Poisson--Lie T--plurality of three--dimensional conformally invariant sigma models II: Nondiagonal metrics and dilaton puzzle}

\author{L. Hlavat\'y \\
Faculty of Nuclear Sciences and Physical Engineering,
\\ Czech Technical University,
\\ B\v rehov\'a 7, 115 19 Prague 1, Czech Republic
\\ \email{hlavaty@fjfi.cvut.cz} }
\author{L. \v Snobl \\
Centre de Recherches Math\'ematiques,  Universit\'e de Montr\'eal, \\
P.O. Box 6128, Centre-ville Station, Montr\'eal (Qu\'ebec), H3C 3J7 Canada \\
{\rm and} \\
Faculty of Nuclear Sciences and Physical Engineering, \\
Czech Technical University, \\
B\v rehov\'a 7, 115 19 Prague 1, Czech Republic \\
\email{Libor.Snobl@fjfi.cvut.cz} }

\abstract{We look for 3--dimensional Poisson--Lie dualizable sigma models that satisfy the vanishing
$\beta$--function equations with constant dilaton field. Using the Poisson--Lie T--plurality we then
construct 3--dimensional sigma models that correspond to various decompositions of Drinfeld double. Models with
nontrivial dilaton field may appear. It turns out that for ``traceless'' dual  algebras they satisfy the
vanishing $\beta$--function equations as well.

In certain cases the dilaton cannot be defined in some of the dual models. We provide an explanation why
this happens  and give criteria predicting when it happens.}

\begin{document}

\section{Introduction}
As it was shown in \cite{klse:dna}, Poisson--Lie T--dual sigma models are given by a Manin triple, i.e. a
decomposition of a classical Drinfeld double, and by an invertible constant matrix $E_0$. Construction of the
dual models is described e.g. in \cite{klse:dna},\cite{kli:pltd} and \cite{tvu:pltdpi}. Examples of dual models
are given e.g.in \cite{vall:su2},\cite{sfe:pltd},\cite{jare:pltd},\cite{bomo:pltd},\cite{klim:ybsm} and
\cite{kun:tdpl}.

The fact that for a given Drinfeld double several Manin triples may exist leads to the notion of Poisson--Lie
T--plurality \cite{unge:pltp}. In our previous paper \cite{hlasno:3dsm1} we presented the Poisson--Lie
T--plurality for a rather special class of sigma models. We looked for the conformally invariant models with the
diagonal matrix $E_0$ and vanishing dilaton field. We obtained models for the Manin triples $(6_0|1),(7_0|1)$
(for the notation see Appendix \ref{app}) and then investigated their associated models.

In this paper we follow a more systematic approach, namely, we do not restrict ourselves to the diagonal
$E_0$'s\footnote{This explains the subtitle of the current paper. It is
a bit imprecise since
even diagonal matrices $E_0$ in most cases give nondiagonal metrics. Because $E_0$ coincides with the sum of
metric and torsion potential at the group unit, a completely precise subtitle would be ``metrics nondiagonal at
the group unit''.}. That provide us with much larger set of the three--dimensional conformally invariant sigma
models. We present general forms of the conformally invariant sigma models with vanishing (more precisely,
constant) dilaton field on the Manin triples $(1|1),(2|1),(3|1),(4|1),(5|1),(6_0|1),(7_0|1)$ and using the
Poisson--Lie T--plurality we construct the conformally invariant sigma models on the other Manin triples in the
corresponding Drinfeld doubles.

To set our notation let us briefly review the construction of the Poisson--Lie T--plural sigma models
\cite{unge:pltp} in the next two sections. In Section \ref{dilpuzz} we discuss the problems with transformations
of dilaton field. Obtained three--dimensional conformally invariant sigma models are given in Section
\ref{3dsm}.

\section{Construction of Poisson--Lie T--dual sigma models} The Poisson--Lie T--dual sigma models are constructed by means of  Drinfeld
doubles. The Drinfeld double $D$ is defined as a connected Lie group such that its Lie algebra $\cd$ can be
decomposed into a pair of subalgebras $\cg$, $\tcg$ maximally isotropic with respect to a symmetric
ad--invariant nondegenerate bilinear form $\langle\, .,.\rangle $ on $\cd$. The dimensions of the subalgebras
are equal and due to the ad-invariance of $\langle\, .,.\rangle $ the algebraic structure of $\cd$ is determined
by the structure of the maximally isotropic subalgebras (see Appendix \ref{app}).

The Lagrangian of the dualizable sigma models \be {L}= F_{ij}(y)\partial_- y^i\partial_+ y^j,\ \
i,j=1,\ldots,n=\dim\cg \lbl{sigm} \ee can be rewritten in terms of right--invariant fields as\footnote{$G$
$(\tilde G)$ is the subgroup of $D$ whose Lie algebra is $\cg$ $(\tcg)$. {All constructions are in general
permissible only locally, in a vicinity of the group unit.}}
\be {L}=E_{ab}(g)(\partial_- gg^{-1})^a(\partial_+
gg^{-1})^b, \;  g \in G \lbl{rflag} \ee where
 \be
E(g)=(E_0^{-1}+\Pi(g))^{-1},
\ \ \ \Pi(g)=b(g)a(g)^{-1} = -\Pi(g)^t.\lbl{poiss}\ee and $a(g),b(g),d(g)$ are $n\times n$ submatrices of the
adjoint representation of the group $G$ on $\cd$ in the basis $(X_i,\tilde X^j)$ \footnote{ t denotes
transposition.} \be Ad(g)^t  =  \left ( \begin{array}{cc}
  a(g)&0  \\ b(g)&d(g)  \end{array} \right ), \lbl{adg}\ee
\be a(g)^{-1}=d(g)^t,\ \ \ b(g)^t a(g) = -a(g)^t b(g). \ee It means that \be F_{ij}(y)=e_i^a (g(y))E_{ab}(g(y))
e_j^b(g(y))\lbl{metorze} \ee where $e_i^a$ are components of right--invariant forms (vielbeins)
 $e_i^a(g)= \left( ({\rm d}g)_i \  . g^{-1}  \right)^a$ and $y^i$ are local coordinates of $g\in G$.

By a modification of this procedure one can construct dual models even for noninvertible matrices $E_0$ (see
\cite{alkltse:qpl}) but we shall not consider such models here, because then the dual model is not of the form
(\ref{rflag}).

The covariant tensor field $F$ on $G$ is thus determined by the decomposition $\cd=(\cg|\tcg)$ and by the matrix
$E_0$. It can be understood as a sum of {the} metric and {the} torsion potential defining the geometric
properties of the manifold $G$. Necessary condition for invertibility of the metric of sigma models is \be
\det(E_0+E_0^t)\neq 0.\lbl{eott} \ee It turns out that usually this condition is also sufficient.

{The ultraviolate finiteness of the quantum version of the model} is guaranteed by the conformal invariance of
the model. To achieve this invariance at the one--loop level we must add another term containing the so--called
dilaton field to the Lagrangian. The dilaton field  $\Phi$ can be understood as an additional function on $G$
that defines the quantum nonlinear sigma model. The conformal invariance of the model is guaranteed by vanishing
of the so--called $\beta$--function. At the one--loop level the equations for vanishing of the $\beta$--function
read
\begin{eqnarray}
\label{bt1} 0 & = & R_{ij}-\bigtriangledown_i\bigtriangledown_j\Phi- \frac{1}{4}H_{imn}H_j^{mn}, \\
 \label{bt2} 0 & = & \bigtriangledown^k\Phi H_{kij}+\bigtriangledown^k H_{kij},  \\
 \label{bt3} 0 & = & R-2\bigtriangledown_k\bigtriangledown^k\Phi- \bigtriangledown_k\Phi\bigtriangledown^k\Phi- \frac{1}{12}H_{kmn}H^{kmn}
 \end{eqnarray}
where the covariant derivatives $\bigtriangledown_k$, Ricci tensor $R_{ij}$ and Gauss curvature $R$ are
calculated from the metric \be G_{ij}=\frac{1}{2}(F_{ij}+F_{ji}) \ee that is also used for lowering and raising
indices, and the torsion is \be H_{ijk}=\partial_i B_{jk}+\partial_j B_{ki}+\partial_k B_{ij} \lbl{torsion}
\ee where \be B_{ij}=\frac{1}{2}(F_{ij}-F_{ji}). \lbl{torpot}\ee

\section{The Poisson--Lie T--plurality}\lbl{secpltp}
The possibility to decompose some Drinfeld doubles into more than two Manin triples\footnote{Two decompositions
always exist, $(\cg|\tcg)$ and $(\tcg|\cg)$.} enables us to construct more than two equivalent sigma models and
this property is called Poisson--Lie T--plurality \cite{unge:pltp}. Let $\{ X_j,\tilde X^k\},\
j,k\in\{1,...,n\}$ {be} generators of Lie subalgebras {$\cg,\tcg$} of a Manin triple associated with the
Lagrangian \rf{rflag}) and $\{ U_j,\tilde U^k\}$ are generators of some other Manin triple {$(\cg_u,\tcg_u)$} in
the same Drinfeld double related by the $2n\times 2n$ transformation matrix:

\be {\left ( \begin{array}{c}
  \vec X \\ \vec {\tilde X} \end{array} \right )}
  =  \left ( \begin{array}{cc}
 P&Q  \\ R&S  \end{array} \right )\left ( \begin{array}{c} \vec U\\ \vec {\tilde  U} \end{array} \right),
\lbl{trsfmat}\ee where
$$ \vec X=(X_1, \ldots, X_n)^t, \ldots, \; \vec {\tilde  U} =(\tilde U^1,\ldots,\tilde U^n)^t.$$
The transformed model is then given by {the} Lagrangian of the same form as  \rf{rflag}) but with $E(g)$
replaced by
\be E_u(g_u)=M(N+\Pi_u\,M)^{-1}=(E_{0u}^{-1}+\Pi_u)^{-1}\lbl{eg} \ee
where
 \be M=S^tE_0-Q^t, \ \
N=P^t-R^t E_0,\ \  E_{0u}=MN^{-1} \lbl {mn} \ee
 and $\Pi_u$ is calculated by \rf{poiss})
from the adjoint representation of the group $G_u$ generated by $\{U_j\}$. 
The transformation of $E_0$ corresponds to the invariance of 
$${\cal E^{+}}={\rm span} \{ X_j+(E_0)_{jk} \tilde X^k\} ={\rm span} 
\{ U_j+(E_{0u})_{jk} \tilde U^k\}.$$
Note that for $P=S=0,\ Q=R={\bf 1}$
we get the dual model with $E_{0u}=E_0^{-1}$, corresponding to the interchange $\cg \leftrightarrow \tcg$
so that the duality transformation is a special case of the plurality transformation \rf{trsfmat}) -- \rf{mn}).

In quantum theory the duality or plurality transformation must be supplemented by a correction that comes from
integrating out the fields on the dual group $\tilde G$ in path integral formulation. In some cases it can be
absorbed at the 1-loop level into the transformation of the dilaton field $\Phi$ . The transformation of the
tensor $F$ that follows from \rf{eg}) must be accompanied by the transformation of the dilaton \cite{unge:pltp}
\be \Phi_u= \Phi+{\rm ln | Det} (N + \Pi_u M)| - {\rm ln| Det} ({\bf 1} + \Pi E_0)|+{\rm ln | Det}\,a_u|-{\rm ln
|  Det} \, a| \lbl{dualdil}\ee
where $\Pi_u, a_u$ are calculated as in \rf{adg}),\rf{poiss}) but from the
adjoint representation of the group $G_u$. Unfortunately this transformation of the dilaton field cannot be
applied in general. This problem will be discussed in the next section.

Beside the transformations that follow from \rf{trsfmat}) we can rescale the matrix $E_0$ and thus also $F(y)$
by a constant factor. For that one should  notice that although there are nonisomorphic Manin triples whose
commutation relations differ just by an overall multiplication constant $\kappa$ in all the commutators in the
second subalgebra $\tcg$, such Manin triples lead to equivalent models.  The reason is that such rescaling leads
to
$$ a(g) \rightarrow a(g), \; b(g) \rightarrow \kappa\, b(g), \; d(g) \rightarrow d(g) $$
and consequently to rescaling of the metric of such model \be E_0 \rightarrow \frac{E_0}{\kappa}, \;  F
\rightarrow \frac{F}{\kappa}. \lbl{scalee} \ee It is easy to see that such transformation converts one solution
of the vanishing $\beta$--function equations (\ref{bt1})--(\ref{bt3}) into another one, since all terms in each
of the equations scale in the same way. In fact this transformation corresponds to the rescaling of string
tension if the model (\ref{rflag}) is interpreted in the context of string theory.

\section{Dilaton puzzle}\lbl{dilpuzz}
For further reference, it is convenient to introduce in accordance with \cite{unge:pltp}, \be\label{Phi0}
\Phi^{(0)} = \Phi - {\rm ln | Det} ({\bf 1} + \Pi E_0)|-{\rm ln | Det} \, a|. \ee One may introduce also
\be\label{Phi0u} \Phi^{(0)}_u = \Phi_u - {\rm ln | Det} ({\bf 1} + \Pi_u M N^{-1}) |-{\rm ln | Det} \, a_u| =
\Phi^{(0)} + {\rm ln |Det} \, N| \ee showing that although  $\Phi^{(0)}$ can be considered a function on the
whole Drinfeld double by trivial extension \be\label{dilext} \Phi^{(0)} (g.\tilde g) = \Phi^{(0)} (g), \ee it is
not defined unambiguously by the Drinfeld double and the subspace ${\cal E}^{+}$; the Manin triple must be also
specified. Starting from equivalent models on different Manin triples one obtains $\Phi^{(0)}$s differing by an
additive function of {eventual} spectator fields. (In our paper we shall not consider spectators and those
$\Phi^{(0)}$s may differ by a constant.) On the other hand, {from (\ref{Phi0u}) follows} that the dependence of
$\Phi^{(0)}$ on the coordinates of the Drinfeld double is the same for any choice of the Manin triple. This
shall prove to be important in future considerations.

As shown in \cite{unge:pltp,hlasno:3dsm1}, in all known examples of Poisson--Lie T--plural models the
conformal invariance is preserved under the T--plurality transformations (\ref{eg}), (\ref{dualdil}) provided
the new dilaton $\Phi_u$ is well defined. That it may not be the case was observed in \cite{unge:pltp} in an
example with a spectator field present and in this paper we shall present
also other examples without spectators. In both cases new dilaton cannot be defined because it depends on the coordinates on the
subgroup ${\tilde G}_u$ which were presumably integrated out in path integral. The reason for this  was not
understood. In the following we shall give a criterion when the new dilaton $\Phi_u$ exists and also
an explanation from where the trouble arises.

An obvious necessary and sufficient condition for the existence of well--defined $\Phi_u$ is that $\Phi_u$
doesn't depend on the coordinates on the subgroup ${\tilde G}_u$. In the light of the definitions (\ref{Phi0}),
(\ref{Phi0u}) of $\Phi^{(0)}$ and $\Phi^{(0)}_u$ this is equivalent to the condition that $\Phi^{(0)}_u$ doesn't
depend on the element $\tilde g_u$ of ${\tilde G}_u$ in the decomposition\footnote{recall (\ref{dilext})}
$$l=g_u .\tilde g_u, \; g_u \in G_u, \; \tilde g_u \in \tilde G_u$$
and consequently also that $\Phi^{(0)}$ doesn't depend on $\tilde g_u$. Introducing in a vicinity of the group
unit a parametrization of elements of the subgroup ${\tilde G}_u$ by an exponential\footnote{and recalling that
right action corresponds to left--invariant vector fields} we arrive at the following necessary and sufficient
criterion for the existence of the new dilaton $\Phi_u$.

\begin{theorem}\label{thm1}
The dilaton (\ref{dualdil}) for the model defined on the group $G_u$ exists if and only if
$$ \tilde U \Phi^{(0)} (g.\tilde g)=
\frac{ {\rm d}}{{\rm d} t} \Phi^{(0)} \left( g.\tilde g. \ {\rm exp}(t \tilde U) \right) |_{t=0} =0, \; \forall
g \in G_u, \; \forall \tilde g \in \tilde G_u, \; \forall \tilde U \in \tcg_u,   $$ where $\tilde U \in \tcg_u$
is extended as a left--invariant vector field on $D$.
\end{theorem}

This criterion has an obvious computational disadvantage -- although one can rather easily express $U, \tilde U$
in terms of basis elements $X_j,\tilde X^j$ of $\cg, \tcg$ (the transformation between subalgebras is known, see
(\ref{trsfmat})), we also need to rewrite $g_u . \tilde g_u(t), \ g_u \in G_u, \tilde g_u(t) \in \tilde G_u$ as
$h(t).\tilde h(t), \ h(t) \in G, \tilde h(t) \in \tilde G$. Working in the vicinity of the group unit $e$
this can be accomplished using representation of group elements by exponentials and employing
Campbell--Baker--Hausdorff formula but such computation can be quite involved. For applications it seems to be
much easier to use a weaker necessary condition.
\begin{theorem}\label{thm2}
A necessary condition for the existence of the dilaton (\ref{dualdil}) for the model defined on the group $G_u$
is
$$ \tilde U \Phi^{(0)} ( e )
= \frac{ {\rm d}}{{\rm d} t} \Phi^{(0)} ({\rm exp}(t \tilde U) )|_{t=0} =0, \; \forall \tilde U \in \tcg_u.$$
\end{theorem}
In practice it means to check the matrix $R$ from (\ref{trsfmat}) because\footnote{{ Note that the second
equality follows from the fact that transformation (\ref{trsfmat}) preserves the bilinear form $\langle\,
.,.\rangle $, i.e. the relations (\ref{brackets}).}} \be
 \left ( \begin{array}{c} \vec U\\ \vec{\tilde U} \end{array} \right )=\left ( \begin{array}{cc}
 P&Q  \\ R&S  \end{array} \right )^{-1}\left ( \begin{array}{c} \vec X\\ \vec{\tilde X} \end{array} \right )=
 \left ( \begin{array}{cc}
 S^t& Q^t  \\ R^t& P^t  \end{array} \right )
 \left ( \begin{array}{c} \vec X\\ \vec{\tilde X} \end{array} \right )\label{rtilda}\ee
so that
\be \tilde U^k \Phi^{(0)} ( e )= R^{jk} X_j\Phi^{(0)} (e) =R^{jk}\frac{\partial
\Phi^{(0)}(y)}{\partial y_j}|_{y=0}\label{checkthm2} \ee
due to the convention (\ref{parg}).

 In all examples known to us the necessary condition
proved to be also sufficient -- the new dilaton is in such cases constructed explicitly. Special cases of
matrices from \rf{rtilda}) can be found in \cite{snohla:ddoubles} but here we work with their general
forms.

It is rather surprising that in all further examples it is either possible to find new dilatons for
transformations to all isomorphic Manin triples (i.e. isomorphic subalgebras immersed in different ways into
$\cd$) or it is impossible for all of them. This coincidence seems to indicate some nontrivial relation between
dilatons as solutions of (\ref{bt1}--\ref{bt3}) for {the} metric and {the} torsion given by \rf{rflag}) and the
Poisson--Lie T--plurality transformations.

Also it is clear that if $\Phi^{(0)}$ is a nontrivial (i.e. nonconstant) function on $G$ then it is impossible
to perform Poisson--Lie T--plurality transformation to the dual group $\tilde G$, i.e. to the Manin triple
obtained by just {interchanging} the subalgebras in the original one (since in this case $\tcg_u=\cg$ and exists
$X \in \cg$ such that $X \Phi^{(0)} \neq 0$). This casts additional doubts on the suitability of the original
term ``duality'' in quantum theory.

Another strange discovery is that in some cases when one starts from conformally invariant model and the new
dilaton (\ref{dualdil}) doesn't exist, there is nevertheless a function $\Phi_u'$ on ${G}_u$ such that the
1--loop $\beta$--function is vanishing  also in the new model. Whether it is possible to somehow relate such
models is unclear (new dilatons were in those cases guessed and no relations to the original {ones are} known).
\medskip

Also the origin of the dilaton puzzle now becomes clear. When the transformation of the dilaton was derived in
\cite{unge:pltp} using path integral (see Section 3 therein), the crucial step was integrating out the
dependence on the auxiliary group $\tilde G$. This was performed using a change of variables from elements of
the group $\tilde G$ to components of invariant 1--forms $\tilde g^{-1} \partial \tilde g$ and integrating
functional $\delta$--function. Terms that appeared in the regularization of a certain functional determinant
were then absorbed into  a shift of a ``naive'' or ``bare'' dilaton $\Phi^{(0)}$ giving the relation between
$\Phi^{(0)}$ and the ``true'' dilaton $\Phi$ as given in (\ref{Phi0}). During this computation it was tacitly
assumed that $\Phi^{(0)}$ is not involved in the integration over $\tilde G$, i.e. that it doesn't depend on it.

It becomes clear that if $\Phi^{(0)}$ depends on $\tilde G$ the integration cannot be performed in a similar
manner -- inverting the relation between invariant 1--forms $\tilde g^{-1} \partial \tilde g$ and group elements
$\tilde g$ gives rise to path--ordered exponentials, i.e. nonlocal terms, and the resulting object can be hardly
interpreted as a dilaton. It is an open question whether this obstacle can be somehow circumvented or reinterpreted.

\section{Sigma models on solvable three-dimensional groups}\lbl{3dsm}
It follows from the construction of classical dualizable models that they are given by a Manin triple
$(\cg|\tcg)$ and $n\times n$ invertible matrices $E_0$ that satisfy \rf{eott}). The quantum models include the
dilaton field and require moreover that the equations (\ref{bt1}) -- (\ref{bt3}) hold. They are equations for
the tensor field $F$ (given by the Manin triple and $E_0$) and the dilaton field $\Phi$. We shall solve them for
Manin triples $(\cg|\tcg)$ where $\cg$ are solvable three dimensional Lie algebras without a
parameter\footnote{{Presently we are not able to perform similar analysis for the Bianchi algebras ${\bf
6_a,7_a}$ because of
 computational difficulties.}} in the Bianchi classification (see \cite{bianchi} or e.g. \cite{Landau})
 and $\tcg$ is three dimensional abelian Lie algebra (i.e. for the triples
$(1|1),(2|1),(3|1),(4|1),(5|1),(6_0|1),(7_0|1)$). Moreover we shall assume that $\Phi=const$. Such solutions
exist only for special forms of $E_0$ given below. Afterwards we shall investigate the sigma models that can be
obtained  by the transformation of Drinfeld doubles \rf{trsfmat}). We shall omit the models corresponding to the
Manin triples where the subalgebra $\tcg$ is not traceless, {i.e. $\tilde{f_{ij}}^j\neq0$,}
 because it is known that these quantum sigma models exhibit the so--called gravitational anomaly
 \cite{grv:anomaly}--\cite{aagl:anomaly}. It means that we
will investigate the following Drinfeld doubles and decompositions (for the notation see Appendix \ref{app} or
\cite{snohla:ddoubles}):
\begin {itemize}
\item DD11: $(5|1)$ $\cong$ $(6_0|1)$ $\cong$ $(1|6_0)$ $\cong$ $(5.ii|6_0)$
$\cong$ $(5|2.i)$,
\item DD12: $(4|1)$ $\cong$  $(6_0|2)$ $\cong$ $(2|6_0)$ $\cong$ $(6_0|4.ii)$ $\cong$ $(4|2.i)$ $\cong$ $(4|2.ii)$,
\item DD13: $(3|1)$ $\cong$ $(3|2)$,
\item DD15: $(7_0|1)$ $\cong$ $(1|7_0)$,
\item DD19: $(2|1)$ $\cong$ $(1|2)$.
\item DD22: $(1|1)$.
\end {itemize}

Often we do not display {the} tensor fields $F_{ij}$ because they are usually too complicated (for examples see \cite{hlasno:3dsm1}) but only the matrices $E_0=F(e)$ from which the tensor fields can be reconstructed by
\rf{poiss}) and \rf{metorze}).

As all groups we are dealing with in the following are solvable we can use the parametrization of $g\in G$ in
the form
\be
g(y)=\exp (y_1X_1)\ \exp (y_2X_2)\ \exp (y_3X_3)\ \lbl{parg}
\ee
 where $y_j$ are coordinates {on} the
group manifold and $X_j$ are the group generators whose commutation relations are given in the Appendix
\ref{app}. We shall not distinguish between $y$ corresponding to different groups by further indices, the group
follows from the context. If we need to consider coordinates on different groups in one expression, we denote
coordinates on the second group by $x$. We assume $\epsilon=\pm 1$ in the following.

\subsection{Sigma models on DD11 starting from $(5|1)$}

Constant dilaton exists for {the} decomposition $(5|1)$ and

\be E_0=\left(\matrix{ p&u&v\cr w&q&g \cr z&g&r
 \cr  } \right),\ \  qr= g^2.
   \lbl{e51}\ee
More precisely, the dilaton $\Phi=C\in \real$ and the tensors  \be F_{ij}(y)=\left(\matrix{
p&u\,{\rm e}^{-y_1}&v\,{\rm e}^{-y_1}\cr w\,{\rm e}^{-y_1}&q\,{\rm e}^{-2y_1}&g\,{\rm e}^{-2y_1} \cr
z\,{\rm e}^{-y_1}&g\,{\rm e}^{-2y_1}&r\,{\rm e}^{-2y_1}
 \cr  } \right)
 ,\ \  qr= g^2
   \ee
constructed by \rf{metorze}) from $(5|1)$ and \rf{e51}), satisfy the $\beta$--function equations (\ref{bt1}) --
(\ref{bt3}).

The metrics are invertible if and only if $g(v+z)\neq r(u+w)$ for $r\neq 0$ and $q(v+z)\neq 0$ for $r=0$. The
corresponding sigma models are torsionless (i.e. $H_{ijk}=0$) and their metrics are flat (i.e. their Riemann
{tensors} vanish). \rf{e51}) is the general form of $E_0$ for which a solution of the $\beta$--function
equations with constant dilaton exists.\footnote{All calculations done in this section were performed by virtue
of Mathematica and Maple so that similar propositions depend on their capability to find all solutions. On the
other hand these programs were applied simultaneously and independently so the results seem to be  rather
reliable.}

All models given above can be obtained from those with
\be E_0=\pm\left(\matrix{ 0&0&1\cr 0& 1&0 \cr 1&0&0 \ \cr  } \right)
   \ee by plurality transformations \rf{trsfmat})--(\ref{mn}) that leave the structure constants of
$(5|1)$ invariant and we can get rid off the sign
   by the scaling transformation \rf{scalee}).

Besides that by other plurality transformations we can get models corresponding to {other decompositions} of the
DD11. The dilaton field for these models can be obtained from \rf{dualdil}). However, before using it we should
check whether this formula provides us with a function independent of the coordinates of the subgroup $\tilde
G_u$. As mentioned in Section \ref{dilpuzz} it is rather difficult, nevertheless, we can check at least the
necessary condition given by the Theorem \ref{thm2}. For that we shall need the value of $\Phi^{(0)}$ introduced
in Section \ref{dilpuzz}. From \rf{Phi0}) we get \be \Phi^{(0)}(y)=C-2y_1 \lbl{Phi051}\ee
 so that the
condition that follows from (\ref{checkthm2}) now reads \be -2R^{1k}=0, \ k=1,2,3 \ee
 and by inspection of the transformation matrices we can find if it is
satisfied. The transformation $(y,\tilde y)\rightarrow (x,\tilde x)$ of coordinates on $D$ then can be derived
by decomposing an element of $D$ in different ways, i.e.
\be
  l=g(y)\tilde g(\tilde y)=g_u(x)\tilde g_u(\tilde x).
\lbl{lgg}\ee

Models  different from those above and given by decompositions of the DD11 with the ``traceless'' second factor
correspond to the Manin triples $(5|2.i),\ (6_0|1),\ (1|6_0)$ and $(5.ii|6_0)$. Let us investigate these
possibilities.
\begin{itemize}
\item Decomposition $(5|2.i)$. There are six matrices with up to eight free parameters
transforming $(5|1)$ to $(5|2.i)$ by \rf{trsfmat}). Using \rf{mn}) we find that all of them lead to models given
by \be\label{52imodel}
 \tilde{E_0}=\left(\matrix{ P&U&V\cr Y&Q&-1-\epsilon \cr X&1+\epsilon&0
 \cr  } \right)
   \ee
or \be \tilde{E_0}=\left(\matrix{ P&U&V\cr Y&(G+1+\epsilon)^2/J&G \cr X&G+2\epsilon+2&J
 \cr  } \right).
   \ee
The metrics are invertible if and only if $Q(V+X)\neq 0$ and  $(2\epsilon+2+G)(V+X)\neq J(U+Y)$
respectively. The corresponding sigma model are torsionless and their metrics are flat.

By inspection of the transformation matrices one finds that for all of them $R^{1k}=0$ and from \rf{lgg}) one
gets $y_1=-\epsilon x_1$.
Using \rf{dualdil}) one finally finds
that the transformed dilaton $\Phi_u$ is constant so that the $\beta$-function equations (\ref{bt1})--(\ref{bt3})
are again satisfied.

\item Decomposition $(6_0|1)$. There are two matrices
transforming $(5|1)$ to $(6_0|1)$. They produce models given by \be\label{E610} \tilde{E_0}=\left(\matrix{
Q&2\epsilon Q-Y&V\cr Y& Q&G \cr X&H&J
 \cr  } \right).
   \ee
The metrics are invertible if and only if $Q(G+H-\epsilon(V+X))\neq 0$. The corresponding sigma models
are torsionless but their metric is not flat. Neither Riemann nor Ricci tensors vanish. Gauss curvature is
zero. Similarly as in the previous case one can check that the necessary condition $ -2R^{1k}=0 $ for
application of the dilaton formula \rf{dualdil}) is satisfied and the transformation of the coordinates is
$y_1=-\epsilon x_3$. The $\beta$-function equations (\ref{bt1})--(\ref{bt3}) are satisfied for the dilaton
field \be\Phi(x)=C+2\epsilon x_3\ee obtained from \rf{dualdil}).

\item Decompositions $(1|6_0)$. Transformations (\ref{trsfmat})--(\ref{mn}) give models with
\be \tilde {E_0}=\left(\matrix{ P&U&V \cr Y&Q&G \cr
 X&H&J
 \cr  } \right)\label{e0gen}\ee
where the elements satisfy \be PJ-VX-QJ+GH=0,\ee \be GH-PJ-QJ+VX=\epsilon(HV-UJ+GX-JY).
   \ee
If $J=0$ then
 \be \tilde{E_0}=\left(\matrix{ P&U&\epsilon G\cr Y&Q&G \cr
\epsilon H&H&0
 \cr  } \right)
   \ee
The corresponding sigma models are torsionless and their metrics are flat.

If $J\neq 0$ then
\be\label{E160} \tilde{E_0}=\left(\matrix{ P&[2\epsilon(PJ-VX)+HV+GX]/J-Y&V \cr Y&P+(GH-VX)/J&G \cr
 X&H&J
 \cr  } \right).\ee
The corresponding sigma models have nontrivial torsion and their metrics are not flat. The form of dilaton field
is not known because $R^{13}=\pm 1$ so that the necessary condition for existence of the dilaton
independent of $\tilde x$ is violated and the formula \rf{dualdil}) is not applicable.

\item Decomposition $(5.ii|6_0)$. Transformations (\ref{trsfmat})--(\ref{mn}) give models {with
$\tilde E_0$ given by \rf{e0gen})} where the elements satisfy \be PJ-(V-1)(X+1)-QJ+(G-1)(H+1)=0,\ee
$$2(\epsilon-1)+PJ-(V-1+\epsilon)(X+1-\epsilon)+QJ-$$
\be(G-1+\epsilon)(H+1-\epsilon)+\epsilon(HV-UJ-JY+GX)=0.
   \ee

If $J=0$ then
 \be \tilde {E_0}=\left(\matrix{ P&U&\epsilon (G-1)+1\cr Y&Q&G \cr
\epsilon (X+1)-1&H&0
 \cr  } \right).
   \ee
The corresponding sigma models are torsionless and their metrics are flat.

 If $J\neq 0$ then
 \be Q=P+\frac{G-H+GH-V+X- VX}{J},\ee
 \be U=2\epsilon P-Y+\frac{2(\epsilon-1)+G-H+HV+(2\epsilon-1)(X-V)+GX-2\epsilon VX}{J}.\ee
The corresponding sigma models have nontrivial torsion and their metrics are not flat. The form of the dilaton
field is not known because $R^{13}=\pm 1$ so that the formula \rf{dualdil}) is not applicable.

\end{itemize}

\subsection{Sigma models on DD11 starting from $(6_0|1)$}
Models {on the decomposition\footnote{This set of models was already investigated in \cite{hlasno:3dsm1},
although not all possible forms of $\tilde E_0$ were explicitly presented for each Manin triple. On the other
hand, in \cite{hlasno:3dsm1} formulae for the metric, its Gauss curvature and singularities are usually given in
full detail.} $(6_0|1)$ allow constant dilaton for} \be\label{E601} E_0=\left(\matrix{ p&u&v\cr -u&-p&g \cr
z&h&r
 \cr  } \right)
   \ee
i.e. for the tensor
\be F_{ij}(y)=\left(\matrix{ p & u & v + u\,y_1 + p\,y_2 \cr -u & -p & g - p\,y_1 - u\,y_2
\cr z - u\,y_1 + p\,y_2 & h - p\,y_1 +
   u\,y_2 & r(y) \cr  }\right)\ee
where $$r(y)= r + (g+h)\,y_1 + (v+z)\,y_2+ p\,({y_2}^2-{y_1}^2).$$ The metric is invertible if and only
if $p(4pr+(g+h)^2-(v+z)^2)\neq 0$. The corresponding sigma models are torsionless and their metrics are flat.

The function $\Phi^{(0)}$ is constant in this case so that we have no problems with its possible dependence on
coordinates of auxiliary groups $\tilde G_u$ and the formula \rf{dualdil}) is valid for any decomposition.

All these models can be obtained by a plurality transformation from those with \be E_0=\kappa\left(\matrix{
1&0&0\cr 0&- 1&0\cr 0&0&1
 \cr  } \right)
   \ee
and we can get rid off the overall constant $\kappa\in\real {\setminus\{ 0 \}}$ by the scaling transformation
mentioned {at} the end of Section \ref{secpltp}.

Beside that by the plurality transformations \rf{trsfmat}) we get the sigma models obtained from
\begin{itemize}
\item Decompositions $(1|6_0)$ and
 \be \label{E160601} \tilde{E_0}=\left(\matrix{ P&(HV+GX-JY)/J&V\cr Y&(GH-PJ+VX)/J&G \cr
X&H&J
 \cr  } \right)
   \ee
or
 \be \label{16_0npd}\tilde{E_0}=\left(\matrix{ P&U&-\epsilon G\cr Y&Q&G \cr
\epsilon H&H&0
 \cr  } \right).
   \ee

In the first case the corresponding sigma models have nontrivial torsion and their metrics are not flat. The
dilaton field obtained from (\ref{dualdil}) is \be \Phi(x)=\ln\Big | 1+(G-H)x_1 +(V-X)x_2
+(VX-PJ)(x_1^2-x_2^2)\Big |+C. \ee In the second case the corresponding sigma models are torsionless and their
metrics are flat. The dilaton field obtained from (\ref{dualdil}) is \be \Phi(x)=\ln \Big |(1+G(x_1 -\epsilon
x_2))(1-H(x_1+\epsilon x_2))\Big |+C.\label{16_0npdD}\ee In both cases the $\beta$-function equations
(\ref{bt1})--(\ref{bt3}) are satisfied. Note that there is another solution of (\ref{bt1})--(\ref{bt3}) in the
case (\ref{16_0npd}), namely $\Phi(x)=const.$ while $\Phi$ given by (\ref{16_0npdD}) is a nontrivial solution of
\be\label{trivdil} \bigtriangledown_i\bigtriangledown_j\Phi =0, \; {\bigtriangledown^j \Phi
\bigtriangledown_j\Phi =0,} \; \forall i,j. \ee

\item Decomposition $(5|2.i)$ and
\be \tilde{E_0}=\left(\matrix{ P&U&V\cr Y&Q&\epsilon-1 \cr X&\epsilon+1&0
 \cr  } \right)
   \ee
or \be \tilde{E_0}=\left(\matrix{ P&U&V\cr Y&(G+2)G/J&G \cr X&G+2&J
 \cr  } \right).
   \ee
The corresponding sigma model are torsionless and their metrics are flat. The dilaton field given by
\rf{dualdil}) is constant.
\item Decomposition $(5.ii|6_0)$. Transformations (\ref{trsfmat}) -- (\ref{mn}) give models with
 \be \tilde {E_0}=\left(\matrix{ P&U&V\cr Y&Q&\epsilon(V-1)+1 \cr
X&-\epsilon (X+1)-1&0
 \cr  } \label{e5ii60} \right)
   \ee
or \be \tilde {E_0}=\left(\matrix{ P&\frac{-2 + G - H + V + H V - X + G X - J Y}{J}&V\cr Y&\frac{-2 + G - H + G
H - P J + V - X + V X}{J}&G \cr X&H&J
 \cr  } \right).
   \ee
In the former case the corresponding sigma models are torsionless and their metrics are flat. The dilaton
obtained from \rf{dualdil}) is
$$ \Phi(x)=C+\ln|V{\rm e}^{\epsilon(x_1+x_2)}+(V-1)[{\rm e}^{-x_2}(\epsilon-1)-\epsilon]|$$
\be +\ln|X{\rm e}^{-\epsilon(x_1+x_2)}-(X+1)[{\rm e}^{-x_2}(\epsilon+1)-\epsilon]|\label{dil5ii60}\ee and the
$\beta$-function equations (\ref{bt1})--(\ref{bt3}) are satisfied. Note that once again there is another
solution of (\ref{bt1})--(\ref{bt3}) in this case, namely $\Phi(x)=const.$

In the latter case the corresponding sigma models have nontrivial torsion and their metrics are not flat. The
dilaton obtained from \rf{dualdil})
\begin{eqnarray}
\ln \Big|
\frac{{\rm e}^{-x_1  - x_2 }}{2\,J} \,\left( -2 + G - H + 2\,P\,J - V + X - 2\,V\,X \right.  & + & \nonumber  \\
 {\rm e}^{x_1  + x_2 }\,\left( 2 - 2\,V + 2\,X \right)
 + {\rm e}^{2\,\left( x_1  + x_2  \right) }\,\left( -2 \right. + \left. G - H - 2\,P\,J \right. & + & \nonumber  \\
\left. V - X + 2\,V\,X \right)- 2\,{\rm e}^{2\,x_1  + x_2 }\,\left( -2 + G - H - 2\,P\,J + V \right. & - & \nonumber \\
\left. \left. X  +  2\,V\,X \right)  - 4\,{\rm e}^{x_1 }\,\left( P\,J - \left( -1 + V \right) \,\left( 1 + X \right)  \right)
\right) \Big| +C
& &
\end{eqnarray}
is quite complicated and we are not able to check whether the $\beta$-function equations
(\ref{bt1})--(\ref{bt3}) are satisfied in general. An example of such model was given in \cite{hlasno:3dsm1} and
its $\beta$--function is known to vanish on {the} 1--loop level.

\item Decomposition $(5|1)$. There is no sigma model corresponding to this decomposition because  $\Pi_u(g)=0$
in \rf{eg}) and the matrix $N$ is not invertible for any transformation matrix \rf{trsfmat}).
\end{itemize}
\subsection{Sigma models on DD12 starting from $(4|1)$}
Constant dilaton exists for the decomposition $(4|1)$ and \be E_0=\left(\matrix{ p&u&v\cr w&q&0 \cr z&0&0
 \cr  } \right)
   \ee
i.e for the tensor  \be F_{ij}(y)=\left(\matrix{ p & (u + v\,y_1)\,{{\rm e}^{-y_1}} & v\,{{\rm e}^{-y_1}} \cr (w +
z\,y_1)\,{{\rm e}^{-y_1}} & {q}\,
   {{\rm e}^{-2\,y_1}} & 0 \cr {z}\,{{\rm e}^{-y_1}} & 0 & 0 \cr  }\right).\lbl{e41}\ee
   The metric is invertible if and only if $q(v+z)\neq 0$. The corresponding sigma models are torsionless, their metrics
are flat and $\Phi^{(0)}(y)=C-2y_1$.

All these models can be obtained by a plurality transformation from those with \be E_0=\kappa\left(\matrix{
0&0&1\cr 0&1&0 \cr 1&0&0
 \cr  } \right).
   \ee Once again we can get rid off the overall constant $\kappa\in\real {\setminus\{ 0 \}}$
   by the scaling transformation \rf{scalee}).

Beside that by plurality transformations we get models with
\begin{itemize}
\item Decompositions $(4|2.i)$, $(4|2.ii)$ and
\be\label{42i1} \tilde{E_0}=\left(\matrix{ P&U&V\cr Y&Q&0 \cr X&0&0
 \cr  } \right)
   \ee
or \be \tilde{E_0}=\left(\matrix{ P&U&V\cr Y&Q&-2\epsilon \cr
X&2\epsilon&0
 \cr  } \right)
   \ee
where $\epsilon=1$ for $(4|2.i)$ and $\epsilon=-1$ for $(4|2.ii)$. The metric is invertible if and only if
$Q(V+X)\neq 0$. The corresponding sigma models are torsionless and their metric is flat. The necessary condition
for the dilaton transformation is satisfied and the resulting dilaton is constant.

\item Decomposition $(2|6_0)$ and
\be \tilde{E_0}=\left(\matrix{ P&U&V\cr Y&Q&\epsilon V \cr
X&\epsilon X&0
 \cr  } \right).
   \ee
The metric is invertible only if $[P+Q-\epsilon(U+Y)](V+X)\neq 0$. The corresponding sigma models are
torsionless and their metric is flat so that this model allows constant dilaton in spite of the fact that the
necessary condition for the transformation of (constant) dilaton of $(4|1)$ is violated because $ R^{13}=\pm 1$.
\item Decomposition $(6_0|2)$ and
 \be \tilde{E_0}=\left(\matrix{ P&\epsilon P&V\cr \epsilon P&P&G \cr X&H&J
 \cr  } \right).
   \ee
The metric is invertible if and only if $P[G+H-\epsilon(V+X)]\neq 0$. The corresponding sigma models are
torsionless but their metric is not flat. Neither Riemann nor Ricci tensors vanish. Gauss curvature vanishes.
The necessary condition for the dilaton transformation is satisfied. The dilaton is
$$\Phi(x)=2\epsilon x_3+C$$ and the $\beta$-function equations (\ref{bt1})--(\ref{bt3}) hold.

\item Decomposition $(4.ii|6_0)$ and
\be \tilde{E_0}=\left(\matrix{ P&U&V\cr Y&Q&V \cr X& X&0
 \cr  } \right)
   \ee
or \be \tilde{E_0}=\left(\matrix{ P&U&1+W\cr Y&Q&1-W \cr -1+Z&
-1-Z&0
 \cr  } \right)
   \ee
The corresponding sigma models are torsionless and their metrics are flat so that this is once again a model
allowing a constant dilaton in spite of the fact that the necessary condition for the transformation of
(constant) dilaton of ${(4|1)}$ is violated because $ R^{11}=R^{12}=\pm 1$.

\end{itemize}
\subsection{Sigma models on DD13 starting from $(3|1)$}
Constant dilaton exists for the decomposition $(3|1)$ and \be E_0=\left(\matrix{ p&u&v\cr w&q&-q \cr z&-q&q
 \cr  } \right)
   \ee
i.e. for the tensor \be
F_{ij}(y)=\left(\matrix{ p&\frac{u-v+(u+v){\rm e}^{-2y_1}}{2}&\frac{-u+v+(u+v){\rm e}^{-2y_1}}{2} \cr
\frac{w-z+(w+z){\rm e}^{-2y_1}}{2}&q&-q \cr \frac{-w+z+(w+z){\rm e}^{-2y_1}}{2}&-q&q
 \cr  } \right). \ee
The metric is invertible if and only if $q(u+v+w+z)\neq 0$. The corresponding sigma models are torsionless,
their metrics are flat.

All these models can be obtained by a plurality transformation from those  with \be E_0=\left(\matrix{
0&0&1\cr 0&\epsilon&-\epsilon \cr 1&-\epsilon&\epsilon
 \cr  } \right).
   \ee

Beside that by plurality transformations we get models with
\begin{itemize}
\item Decompositions $(3|2)$ and
\be \tilde{E_0}=\left(\matrix{ P&U&V\cr Y&J&-J \cr X&-J&J
 \cr  } \right)
   \ee
or \be \tilde{E_0}=\left(\matrix{ P&U&V\cr Y&J&-2+J \cr X&2+J&J
 \cr  } \right)
   \ee
where ${\rm sgn}J=\epsilon$. The metric is invertible if and only if $J(U+V+X+Y)\neq 0$ in the first case
and if and only if $J(U-V-X+Y)\neq 0$ in the second. The corresponding sigma models are torsionless and their
metrics are flat. The necessary condition for the dilaton transformation is satisfied and the dilaton is
constant.

\end{itemize}
\subsection{Sigma models on DD15 starting from $(7_0|1)$}
Constant dilaton exists for the decomposition\footnote{This set of models was also investigated in
\cite{hlasno:3dsm1}.}
 $(7_0|1)$ and \be\label{E701} E_0=\left(\matrix{ p&u&v\cr -u&p&g \cr z&h&r
 \cr  } \right)
   \ee
i.e for the tensor  \be F_{ij}(y)=\left(\matrix{ p & u & v - u\,y_1 + p\,y_2 \cr -u & p & g - p\,y_1 - u\,y_2
\cr z + u\,y_1 + p\,y_2 & h - p\,y_1 +
   u\,y_2 &r(y) \cr  } \right)
   \ee
where $$ r(y)= r - (g+h)\,y_1 +  (v+z)\,y_2 +p\,({y_1}^2  +{y_2}^2).$$ The metric is invertible if and only if
$p(4pr-(g+h)^2-(v+z)^2)\neq 0$. The corresponding sigma models are torsionless and their metrics are flat.

The function $\Phi^{(0)}$ is constant in this case so that we have no problems with its possible dependence on
coordinates of auxiliary groups $\tilde G_u$ and the formula \rf{dualdil}) is valid for any decomposition.

All these models can be obtained by a plurality transformation and scaling \rf{scalee}) from those with
\be\label{E701a} E_0=\left(\matrix{ 1&0&0\cr 0&1&0\cr 0&0&\pm 1
 \cr  } \right)
.
   \ee

Beside that by plurality transformations we get
\begin{itemize}
\item Decompositions $(1|7_0)$ and
 \be\label{E170} \tilde{E_0}=\left(\matrix{ (QJ-GH+VX)/J&(HV+GX-JY)/J&V\cr Y&Q&G \cr
X&H&J
 \cr  } \right),
   \ee
where models with definite metric are obtained by plurality from (\ref{E701a}) with $+$ sign, models with indefinite metric from
(\ref{E701a}) with $-$ sign.

The corresponding sigma models have nontrivial  torsion and their metrics are not flat. The dilaton field is \be
\Phi(x)=\ln | 1+(G-H)x_1 +(X-V)x_2 +(QJ-GH)(x_1^2+x_2^2) |+C \ee
 and the $\beta$-function equations (\ref{bt1})--(\ref{bt3}) are satisfied.

\end{itemize}
\subsection{Sigma models on DD19 starting from $(2|1)$}
Constant dilaton exists for the decomposition $(2|1)$ and \be E_0=\left(\matrix{ 0&u&v\cr w&q&g \cr z&h&r
 \cr  } \right)
   \ee
i.e for the tensor  \be F_{ij}(y)=\left(\matrix{ 0&u&v\cr w&q&g+w\,y_2 \cr z&h+u\,y_2 &r+(v+z)\,y_2
 \cr  } \right).
   \ee
The corresponding sigma models are torsionless and their metrics are flat, $\Phi^{(0)}=C.$

All these models can be obtained by a plurality transformation from those with \be E_0=\pm\left(\matrix{
0&0&1\cr 0&1&0 \cr 1&0&0
 \cr  } \right).
   \ee

Beside that by the plurality transformations we get
\begin{itemize}
\item Decompositions $(1|2)$ and
\be\label{E12} \tilde{E_0}=\left(\matrix{ P&U&V\cr Y&Q&G \cr X&H&J
 \cr  } \right),\ \ QJ=GH.
   \ee
For $G= H$ the corresponding sigma models are flat and torsionless, for $G\neq H$ they have nontrivial torsion
and their Gauss curvature is not zero, i.e. their metrics are not flat.

The dilaton field \be \Phi(x) = \ln |1+(G-H)x_1 |+C \ee obtained from \rf{dualdil}) satisfies the vanishing
$\beta$--function equations (\ref{bt1})--(\ref{bt3}).

\end{itemize}
\subsection{Sigma models on DD22}
DD22 denotes the abelian Drinfeld double having just one {class of isomorphic} Manin triples $(1|1)$. Constant
dilaton exists for arbitrary $E_0$. The tensor $F$ is constant so that all models are flat and
torsionless and
can be obtained by a plurality transformation from that with
\be\label{abelmet} E_0=\left(\matrix{ 1&0&0\cr 0&1&0\cr 0&0&\pm 1
 \cr  } \right).
   \ee
\section{Conclusions}

In the present paper we have provided an explanation of the origin of the dilaton puzzle and given criteria
establishing when the new dilaton exists. It became clear that in generic case of $\Phi^{(0)}$ nonconstant (or
in general, not a function of spectator fields only) the simplest dual, obtained by the interchange of the
subalgebras, doesn't exist. This casts some doubts on the use of name ``duality''.

We have presented several sets of mutually equivalent Poisson--Lie T--plural models and
found several
examples of properties not encountered before in the context of Poisson--Lie T--plural models.

Firstly, we know that there are models, namely (\ref{16_0npd}), (\ref{e5ii60}) that allow two different
dilatons, i.e. (\ref{16_0npdD}),{(\ref{dil5ii60}) respectively, and the constant one}. It is quite surprising to
have two seemingly distinct solutions of the vanishing $\beta$--function equations (\ref{bt1}--\ref{bt3}) for
the same flat metric. The only explanation {of (\ref{trivdil})} is that in ``flat'' coordinates the dilaton is a
linear function of coordinates such that its derivative is a null--vector. This knowledge can be helpful if one
attempts to find the flat coordinates.

Secondly, in (\ref{E160601}) and (\ref{16_0npd}) (and similarly in (\ref{E12})) we have an example of {
T--plural} models on the same Manin triple with different matrices $E_0$\footnote{i.e. on isomorphic Manin
triples with the same subspaces ${\cal E}^\pm$, see \cite{hlasno:3dsm1}}, some of them being flat with constant
dilaton, the others being curved and with nontrivial dilaton.

Thirdly, by different choices of $E_0$ one can find models {inequivalent in T--plurality sense} on the same
Manin triple with rather different properties -- compare e.g. (\ref{E610}) and (\ref{E601}). Of course, in
general it is highly unprobable that a dilaton satisfying (\ref{bt1})--(\ref{bt3}) exists for a generic choice
of $E_0$, e.g. general forms (\ref{E610}) and (\ref{E601}) of  $E_0$ are  quite special, fixing in both cases
two relations between elements of $E_0$.

Also it became clear that coordinates based on 1--parameter subgroups which were used in the paper are suitable
for description of invariant vector fields, plurality transformation etc., but may be rather inconvenient for
understanding geometric properties of the models found. On the other hand the group structure may become rather
complicated in coordinates in which geometric properties are transparent, e.g. coordinates such that metric
becomes Minkowski (resp. Euclidean) in flat cases. We know only two cases when a suitable choice of 1--parameter
subgroups leads to explicitly flat coordinates -- namely by renaming the basis elements
$X_1'=X_3,X_2'=X_2,X_3'=X_1$ and simple linear transformation of coordinates one gets explicitly constant
diagonal metric in (\ref{E601}) and (\ref{E701}) for $g=h=v=z=0$. Since our current goal was to get better
understanding of Poisson--Lie T--plurality through investigation of examples we presented the results in
coordinates suited for description of group properties, i.e. via 1--parameter subgroups. In future we plan also
to investigate the geometric structure of the models we found, e.g. whether their metric can be simplified in
some suitable coordinates or whether they can be related to some models with already known properties. {Also it
might easily happen that models with different $E_0$ or even with different group structure can be, as far as
metric, torsion potential and dilaton are considered, transformed one into the other by change of coordinates.
This is the case e.g. in all flat torsion-free, constant dilaton cases with indefinite metric -- they are all
(locally) equivalent to the model with Minkowski metric\footnote{{ Note that metrics (\ref{E701}) and
(\ref{abelmet}) allow also non--equivalent Euclidean signature, which is probably physically less significant.
}}.  It is an open question between which non--flat models a similar identification is also possible. On the other
hand, the group structures involved may have importance in global aspects which are currently not understood in
Poisson--Lie T--plurality setting. One should be also aware that using one fixed transformation (\ref{trsfmat})
one may arrive to different models starting from equivalent, e.g. flat, models with different $E_0$ (as
illustrated e.g. in (\ref{E12})).}

This investigation would be also related to another question of great practical importance. At present, there is
no way of determining whether a given model, i.e. metric together with torsion potential and dilaton prescribed
in some coordinates, can be dualized (pluralized). There exist a criterion given in \cite{klse:dna}, the
so--called generalized isometry condition \be ( {\cal L}_{ v_{c} } F)_{ij}= \tilde{f}^{ab}_c
v_a^mv_b^nF_{im}F_{nj} \lbl{kse}\ee but it assumes that the group structure is already known, i.e. that
left--invariant vector fields $v_a^m$ are given. In the contrary to ordinary isometries with
$\tilde{f}^{ab}_c=0$, generalized isometries satisfying (\ref{kse}) cannot be investigated one by one, they must
be considered as a set due to their nontrivial interplay on the right--hand side of (\ref{kse}). {At the present
there is} no algorithm that would indicate which group one should consider and how it should be expressed in
initially given coordinates. The best algorithm available at the moment seems to be to find first the algebra
of ordinary isometries, i.e. Killing vectors, and the corresponding group, then to determine its subgroups
acting freely on the target manifold. For each of those subgroups $G$ one should then parametrize their orbits
by spectator coordinates (if the action is not transitive) and to determine the Drinfeld double containing Manin
triple $(\cg|1)$. Then one can investigate models given by other decomposition of that Drinfeld double provided
new metric (\ref{eg}) and new dilaton (\ref{dualdil}) exists as explained in Section \ref{dilpuzz} and
\cite{hlasno:3dsm1}. We hope that in future methods or clues for determination of suitable algebras of
generalized isometries of a given model will be found and that it will be possible to find truly Poisson--Lie
T--plural models even for physically interesting metrics with few or without ordinary isometries.

As mentioned before, it is rather surprising that in all known examples it is simultaneously either possible or
impossible to find new dilatons in all models obtained by moving subalgebras $\cg,\tcg$ in the Drinfeld double
without altering their structures (i.e. transformations to isomorphic Manin triples);  such transformation can
be also interpreted as a certain change of the matrix $E_0$ in the given model (\ref{rflag}). This coincidence
seems to indicate a deeper relation between conformal invariance and the Poisson--Lie T--plurality and provides
additional motivation for its further study.

The curvatures of the models defined by (\ref{E160}), (\ref{E170}) and  (\ref{E12}) diverge\footnote{In the
cases (\ref{E160}), (\ref{E170}) this was observed already in \cite{hlasno:3dsm1}.} on hypersurfaces where the
corresponding dilatons are also divergent (i.e. behave like $\ln(0)$). The singular hypersurfaces are
parametrized in coordinate space as hyperbolic cylinder, elliptic cylinder and hyperplane respectively.
Nevertheless, the metric, torsion potential and dilaton appear to have a reasonable continuation behind the
singularity since all of them are well--defined there by formulae (\ref{eg}),(\ref{dualdil}). We don't know at
the moment whether such backgrounds have meaningful physical interpretation, e.g. as branes.

\acknowledgments{ Support of the Ministry of Education of Czech Republic under the research plan MSM210000018 is
gratefully acknowledged. L. \v Snobl thanks Pavel Winternitz, Alfred Michel Grundland and Centre de Recherches
Math\'ematiques for the support of his postdoctoral stay at Universit\'e de Montr\'eal. We are also grateful to
Ji\v r\'{\i} Novotn\'y for a useful discussion about evaluation of path integrals.
}

\appendix
\section {Drinfeld doubles}\label{app}
The Drinfeld double $D$ is defined as a connected Lie group such that its Lie algebra $\cd$ equipped by a
symmetric ad--invariant nondegenerate bilinear form $\langle\, .,.\rangle $ can be decomposed into a pair of
subalgebras $\cg$, $\tcg$ maximally isotropic with respect to $\langle\, .,.\rangle $. The dimensions of the
subalgebras are equal and bases $\{X_i\}, \{\tilde X^i\}$ in the subalgebras can be chosen so that \be \langle
X_i,X_j\rangle =0,\  \langle X_i,\tilde X^j\rangle =\langle \tilde X^j,X_i\rangle =\delta_i^j,\ \langle \tilde
X^i,\tilde X^j\rangle =0. \lbl{brackets}\ee {We shall assume that any basis of Manin triple considered in this
paper satisfies (\ref{brackets}).}

Due to the ad-invariance of $\langle\, .,.\rangle $ the structure
constants of $\cd$ are determined by the structure of its maximally isotropic subalgebras $\cg,\tcg$, i.e. if in
bases $\{X_i\}, \{\tilde X^i\}$ the Lie products are given by
\[ [X_i,X_j]={f_{ij}}^k X_k,\ [\tilde X^i,\tilde X^j]={\tilde {f^{ij}}_k} \tilde X^k\]
then
\be
[X_i,\tilde X^j]={f_{ki}}^j \tilde X^k +{\tilde {f^{jk}}_i} X_k.
\lbl{liebd}\ee
For given Drinfeld double
several Manin triples may exist, i.e.
$${(\cg|\tcg)\cong(\tcg|\cg)\cong(\cg'|\tcg')\cong\ldots}$$
Examples of transformations between $(\cg|\tcg)$ and $(\cg'|\tcg')$ are given in \cite{snohla:ddoubles}. Their
general forms are too extensive to display, nevertheless, they were used throughout this paper.

Classification of real six--dimensional Drinfeld doubles and their decompositions into nonisomorphic Manin
triples are given in \cite{snohla:ddoubles}. Here we shall present only those occuring in this paper. Since only
the subalgebras denoted in Bianchi classification (see \cite{bianchi} or \cite{Landau})
by ${\bf 9,8,7_0,6_0,2,1}$ are traceless, we present in
each Drinfeld double only Manin triples where at least one of the components has the structure of ${\bf
7_0,6_0,2,1}$.

\subsection{Structure of DD11}
$(6_0|1)$ $\cong$ $(6_0|5.ii)$ $\cong$ $(5|1)$ $\cong$ $(5|2.i)$,
  and dual Manin triples ${\bf ( \cg \leftrightarrow \tcg ) } $.
\vskip6mm
\begin{description}
\item[${\bf ( 6_0|1) } $] : \\
$  [X_1,X_2]=0, \; [X_2,X_3] = X_1, \; [X_3,X_1] = -  X_2, $
\\
$ [\tilde{X}^1,\tilde{X}^2]= 0, \, [\tilde{X}^2,\tilde{X}^3] = 0 , \; [\tilde{X}^3,\tilde{X}^1] = 0. $
\vskip4mm  \item[${\bf ( 6_0|5.ii) } $] : \\
$  [X_1,X_2]=0, \; [X_2,X_3] = X_1, \; [X_3,X_1] = -  X_2, $
\\
$ [\tilde{X}^1,\tilde{X}^2]=  - \tilde{X}^1+ \tilde{X}^2 , \, [\tilde{X}^2,\tilde{X}^3] =  \tilde{X}^3, \;
[\tilde{X}^3,\tilde{X}^1] = - \tilde{X}^3. $
\vskip4mm  \item[${\bf ( 5|1) } $] : \\
$  [X_1,X_2]=-X_2, \; [X_2,X_3] = 0, \; [X_3,X_1] = X_3, $
\\
$ [\tilde{X}^1,\tilde{X}^2]= 0, \, [\tilde{X}^2,\tilde{X}^3] = 0 , \; [\tilde{X}^3,\tilde{X}^1] = 0. $
\vskip4mm  \item[${\bf ( 5|2.i) } $] : \\
$  [X_1,X_2]=-X_2, \; [X_2,X_3] = 0, \; [X_3,X_1] = X_3, $
\\
$  [\tilde{X}^1,\tilde{X}^2]= 0, \, [\tilde{X}^2,\tilde{X}^3] = \tilde{X}^1 , \; [\tilde{X}^3,\tilde{X}^1] =  0
.$
\end{description}
\subsection{Structure of DD12}
$(6_0|2)$ $\cong$ $(6_0|4.ii)$ $\cong$ $(4|1)$ $\cong$ $(4|2.i)$ $\cong$ $(4|2.ii)$, and dual Manin triples.
\begin{description}
\vskip4mm  \item[${\bf ( 6_0|2) } $] : \\
$  [X_1,X_2]=0, \; [X_2,X_3] = X_1, \; [X_3,X_1] = -  X_2, $
\\
 $  [\tilde{X}^1,\tilde{X}^2]= \tilde{X}^3, \,
[\tilde{X}^2,\tilde{X}^3] = 0 , \; [\tilde{X}^3,\tilde{X}^1] = 0 .$
\vskip4mm  \item[${\bf ( 6_0|4.ii) } $] : \\
$  [X_1,X_2]=0, \; [X_2,X_3] = X_1, \; [X_3,X_1] = -  X_2, $
\\
$ [\tilde{X}^1,\tilde{X}^2]=  ( - \tilde{X}^1 + \tilde{X}^2 + \tilde{X}^3), \, [\tilde{X}^2,\tilde{X}^3] =
\tilde{X}^3, \; [\tilde{X}^3,\tilde{X}^1] =  - \tilde{X}^3 . $
\vskip4mm  \item[${\bf ( 4|1) } $] : \\
$  [X_1,X_2]=-X_2+X_3, \; [X_2,X_3] = 0, \; [X_3,X_1] = X_3, $
\\
$ [\tilde{X}^1,\tilde{X}^2]= 0, \, [\tilde{X}^2,\tilde{X}^3] = 0 , \; [\tilde{X}^3,\tilde{X}^1] = 0. $
\vskip4mm  \item[${\bf ( 4|2.i) } $] : \\
$  [X_1,X_2]=-X_2+X_3, \; [X_2,X_3] = 0, \; [X_3,X_1] = X_3, $
\\
 $ [\tilde{X}^1,\tilde{X}^2]= 0, \,
[\tilde{X}^2,\tilde{X}^3] = \tilde{X}^1 , \; [\tilde{X}^3,\tilde{X}^1] =  0  .$
\vskip4mm  \item[${\bf ( 4|2.ii) } $] : \\
$  [X_1,X_2]=-X_2+X_3, \; [X_2,X_3] = 0, \; [X_3,X_1] = X_3, $
\\
 $ [\tilde{X}^1,\tilde{X}^2]= 0, \,
[\tilde{X}^2,\tilde{X}^3] = - \tilde{X}^1 , \; [\tilde{X}^3,\tilde{X}^1] =  0  .$
\end{description}
\subsection{Structure of DD13}
$(3|1)$ $\cong$ $(3|2)$ $\cong$ $(3|3.ii)$ $\cong$ $(3|3.iii)$, and dual Manin triples.
\begin{description}
\vskip4mm  \item[${\bf ( 3|1) } $] : \\
$  [X_1,X_2]=-X_2-X_3, \; [X_2,X_3] = 0, \; [X_3,X_1] = X_2+X_3, $
\\
$ [\tilde{X}^1,\tilde{X}^2]= 0, \, [\tilde{X}^2,\tilde{X}^3] = 0 , \; [\tilde{X}^3,\tilde{X}^1] = 0. $
\vskip4mm  \item[${\bf ( 3|2) } $] : \\
$  [X_1,X_2]=-X_2-X_3, \; [X_2,X_3] = 0, \; [X_3,X_1] = X_2+X_3, $
\\
$ [\tilde{X}^1,\tilde{X}^2]=0, \, [\tilde{X}^2,\tilde{X}^3] = \tilde{X}^1 , \; [\tilde{X}^3,\tilde{X}^1] = 0 . $
\end{description}
\subsection{Structure of DD15}
$(7_0|1)$$\cong$ $(1|7_0)$.
\begin{description}
\vskip4mm  \item[${\bf ( 7_0|1) } $] : \\
$  [X_1,X_2]=0, \; [X_2,X_3] = X_1, \; [X_3,X_1] = X_2,
 $
\\
$ [\tilde{X}^1,\tilde{X}^2]= 0, \, [\tilde{X}^2,\tilde{X}^3] = 0 , \; [\tilde{X}^3,\tilde{X}^1] = 0. $
\end{description}
\subsection{Structure of DD19}
$(2|1)$$\cong$ $(1|2)$.
\begin{description}
\vskip4mm  \item[${\bf ( 2|1) } $] : \\
$  [X_1,X_2]=0, \; [X_2,X_3] = X_1, \; [X_3,X_1] = 0,$
\\
$ [\tilde{X}^1,\tilde{X}^2]= 0, \, [\tilde{X}^2,\tilde{X}^3] = 0 , \; [\tilde{X}^3,\tilde{X}^1] = 0. $
\end{description}
\subsection{Structure of DD22}
\begin{description}
\vskip4mm  \item[${\bf ( 1|1) } $] : \\
$  [X_i,X_j]=0,  \; [\tilde{X}^i,\tilde{X}^j] = 0. $
\end{description}


\begin{thebibliography}{10}

\bibitem{klse:dna}
C.~Klim\v{c}\'{\i}k and P.~\v{S}evera, \emph{Dual nonabelian duality
  and the Drinfeld double}, \plb{351}{1995}{455} [\hepth{9502122}].

\bibitem{kli:pltd}
C.~Klim\v{c}\'{\i}k, \emph{Poisson-Lie T-duality}, \npps{46}{1996}{116} [\hepth{9509095}].

\bibitem{tvu:pltdpi}
E.~Tyurin and R.~von Unge, \emph{Poisson-Lie T-duality: the
  path-integral derivation}, \plb{382}{1996}{233} [\hepth{9512025}].

\bibitem{vall:su2}
M.A. Lledo and V.S. Varadarajan, \emph{{\rm SU}(2) Poisson-Lie T-duality}, \lmp{45}{1998}{247}
[\hepth{9803175}].

\bibitem{sfe:pltd}
K.~Sfetsos, \emph{Poisson-Lie T-duality beyond the classical level and
  the renormalization group}, \plb{432}{1998}{365} [\hepth{9803019}].

\bibitem{jare:pltd}
M.A.~Jafarizadeh and A.~Rezaei-Aghdam, \emph{Poisson--{L}ie {T}-duality and {B}ianchi type algebras},
\plb{458}{1999}{470} [\hepth{9903152}].

\bibitem{bomo:pltd}
A.~Bossard and N.~Mohammedi, \emph{Poisson--{L}ie {T}-duality in the string effective action},
\npb{619}{2001}{128} [\hepth{0106211}].

\bibitem{klim:ybsm}
C.~Klim\v{c}\'{\i}k, \emph{Yang--Baxter $\sigma$--models and dS/AdS T-duality}, \jhep{0212}{2002}{051}
[\hepth{0210095}].

\bibitem{kun:tdpl}
K.~E.~Kunze, \emph{T-Duality and Penrose limits of spatially homogeneous and inhomogeneous cosmologies},
\prd{68}{2003}{063517} [\grqc{0303038}].

\bibitem{unge:pltp}
R.~von~Unge, \emph{Poisson-Lie T-plurality}, \jhep{0207}{2002}{014} [\hepth{0205245}].

\bibitem{hlasno:3dsm1}
L.~Hlavat\'y and L.~\v{S}nobl, \emph{Poisson--Lie T--plurality of three--dimensional conformally invariant sigma
models}, \jhep{0405}{2004}{010} [\hepth{0403164}].

\bibitem{snohla:ddoubles}
L.~\v{S}nobl and L.~Hlavat\'y, \emph{Classification of six--dimensional real {D}rinfeld doubles},
\ijmpa{17}{2002}{4043} [\Math{QA}{0202210}].

\bibitem{alkltse:qpl}
A.Yu.~Alekseev, C.~Klim\v c\'{\i}k, and A.A.~Tseytlin, \emph{Quantum Poisson--{L}ie {T}--duality and {W}{Z}{N}{W}
model}, \npb{458}{1996}{430} [\hepth{9509123}]


\bibitem{grv:anomaly}
M.~Gasperini, R.~Ricci and G.~Veneziano, \emph{A problem with non--abelian duality?}, \plb{319}{1993}{438}
[\hepth{9308112}]

\bibitem{gr:anomaly}
A.~Giveon and M.~Ro\v{c}ek, \emph{On non--abelian duality}, \npb{421}{1994}{173} [\hepth{9308154}]

\bibitem{aagl:anomaly}
E.~\'Alvarez, L.~\'Alvarez--Gaum\'e and Y.~Lozano, \emph{On non--abelian
duality}, \npb{424}{1994}{155} [\hepth{9403155}] 

\bibitem{bianchi}
L. Bianchi, \emph{Lezioni sulla teoria dei gruppi continui finite di trasformazioni}, {Enrico Spoerri Editore,}
Pisa, 1918, pp. 550-557.

\bibitem{Landau}
L.D. Landau and E.M. Lifshitz, \emph{The classical theory of fields}, Pergamon Press, London 1987.

\end{thebibliography}
\end{document}